\documentstyle[11pt,cite]{article}

\def\medfigsize{9 truecm}
\def\medcapsize{4.8 truecm}

\def\includefig{y }
\def\dopictures{y }
\ifx\dopictures\includefig
\input epsf.sty			% we use epsf.sty

\else
\newdimen\epsfxsize    		% horizontal size after scaling
\newdimen\epsfysize    		% vertical size after scaling

\def\epsfbox#1{{}}      	% do not print any figure
\fi

\def\tr{{\rm tr}} 		% trace symbol

\def\square{	{\kern1pt\vbox{
	\hrule height 1.2pt\hbox{
	\vrule width 1.2pt\hskip 3pt
   		\vbox{\vskip 6pt}
	\hskip 3pt\vrule width 0.6pt}
	\hrule height 0.6pt}
		\kern1pt}}	% alternative d'Alembertian

\def\np{{\nabla_{\!\!\perp}}}	% nabla_\perp

\newcommand{\mref}[1]{
	(\ref{#1})}		% ref with brackets

\def\donotprint#1{{}}		% if you want to remove something from
\title{
	\begin{flushright}
	{\normalsize TPI--MINN--95--16/T \\
	NUC--MINN--95--14/T \\
	HEP--MINN--95--1346 \\
	hep-ph/9505320\\
	May 1995 \\}
	\end{flushright}
\vglue 2cm
\bf Gluon Production at High Transverse Momentum in the
McLerran-Venugopalan Model of Nuclear Structure Functions
}
\author{
        Alex Kovner, Larry McLerran and Heribert Weigert \\
	{\small\it School of Physics and Astronomy,
	University of Minnesota, Minneapolis, MN 55455}
	}

\date{}

\begin{document}

\maketitle

\begin{center}
{\bf Abstract}\\
\end{center}
We consider the production of high transverse momentum gluons in the
McLerran-Venugopalan model of nuclear structure functions.  We explicitly
compute the high momentum component in this model.  We compute the  nuclear
target size $A$ dependence of the distribution of produced gluons.
\vfill \eject

\section{Introduction}

Understanding parton distributions formed at the initial stages of a
collision between two heavy ions is a very important open problem. In
a previous paper we have set up a formalism for the calculation of the
gluon distribution in the framework of the McLerran-Venugopalan model
of the nuclear structure functions \cite{mv1a,komcwe9502}.  In this
approach the valence quarks in the nuclei are considered as classical
sources of color charge.  This picture is similar to that developed by
Mueller for the gluon structure functions for heavy quark
systems \cite{mu94}. The ``initial values'' of the color distribution
of the valence quarks in the two colliding nuclei is given by
\begin{eqnarray}
 J_{1,2}^\nu(z(x)) = \delta^{\nu \pm}
		\delta (x^\mp) \ g\,\rho_{1,2}(x_\perp)
\end{eqnarray}
As is apparent from the the light cone delta functions
$\delta(x^\mp)$, the quarks in the nuclei appear as infinitely thin
sheets of nuclear matter moving at the speed of light in positive and
negative $z$ directions respectively.

These color charge distributions generate a classical glue field
according to classical Yang-Mills equations
\begin{eqnarray}
\left[D_\mu,F^{\mu\nu}\right] & = &
	\sum\limits_{m=1,2}
	U[A](x,z_m(x))\, J_m^\nu(z_m(x))\,  U[A](z_m(x),x)
\label{eq:fullYM}
\end{eqnarray}
Here $z_{1,2}(x) = \left. x\right\vert_{x^\pm = 0}$ serves as a reference
point used to define the initial value of the charge distribution.
Due to the covariant current conservation $\left[D_\nu,J^\nu(x)\right]
= 0$ this initial distribution evolves along the trajectory of a
particle via parallel transport.  This is the origin of the link
operators
\begin{eqnarray}
U[A](x,z_m(x)) := {\rm P} \exp -ig \int_{z_m(x)}^{x}\!\!
d\omega^\mu_m A^\mu(\omega_m)
\end{eqnarray}
connecting the initial point $z_m(x)$ and the point $x$ on the
trajectory $\omega_m$, which appear on the right hand side of
eq. \mref{eq:fullYM}.

The gluon distribution function is defined in terms of classical
solutions of eq. \mref{eq:fullYM}, and is related to the following
quantity (for precise definition see Section \ref{sec:dist})
\begin{equation}
<A_i(k)A^*_i(k)>_\rho
\end{equation}
Here $A(k)$ is the Fourier transform of the classical solution. The averaging
over the color charge distributions is performed independently for each nucleus
with equal gaussian weights
\begin{equation}
\langle O \rangle_\rho
=\int d\rho_1 d\rho_2\ O\ \exp\left\{-\frac{1}{2\mu^2}\int d^2x_\perp
tr[\rho^2_1(x_\perp)+\rho^2_2(x_\perp)]\right\}
\label{eq:rhoav}
\end{equation}

The purpose of this paper is to calculate this distribution function
perturbatively, to lowest nontrivial order in the inverse powers
of the transverse momentum $\alpha_s\mu/k$.

Basic input in our approach are the classical fields generated by
single nuclei which have been derived earlier in \cite{mv1a,mv1b,mv2}.
These solutions remain valid before the collision and yield initial
conditions for the field after the collision.  The solution in the one
nucleus case is of the form\footnote{On these classical solutions the
link operators in eq. \mref{eq:fullYM} drop out. They may however be
important if one starts to consider quantum corrections in powers of
$\alpha_s$.}
\begin{eqnarray}
A_m^\pm & = & 0
\nonumber \\
A_m^i & = &  \theta(x^\mp) \alpha_{1,2}^i(x_\perp)
\label{onea}
\end{eqnarray}

The functions $\alpha_i$ are implicitly determined  by the ``dimensionally
reduced'' version of the Yang-Mills equations
\begin{eqnarray}
\alpha^i_m & = & -{1 \over{ig}}
	U_m(x_\perp) \partial^i U_m^\dagger (x_\perp)
\nonumber \\
\partial^i \alpha^i_m & = & g\,\rho_m (x_\perp)
\label{eq:sing}
\end{eqnarray}
An obvious property of the solutions eq.(\ref{onea}) is that
the ``transverse'' components of the field
strength vanish $F^{ij} = 0$.  The gauge potentials themselves
vanish
in front of the moving charge and are a pure gauge behind it. The only
physical information is contained in the discontinuity at the worldline
of the quarks which generate the gluon fields. The field strength does
not vanish only on these worldlines and is confined to infinitely
thin sheets.

\begin{figure}[htb]
	\begin{minipage}[b]{\medfigsize} \epsfxsize \medfigsize
		\epsfbox{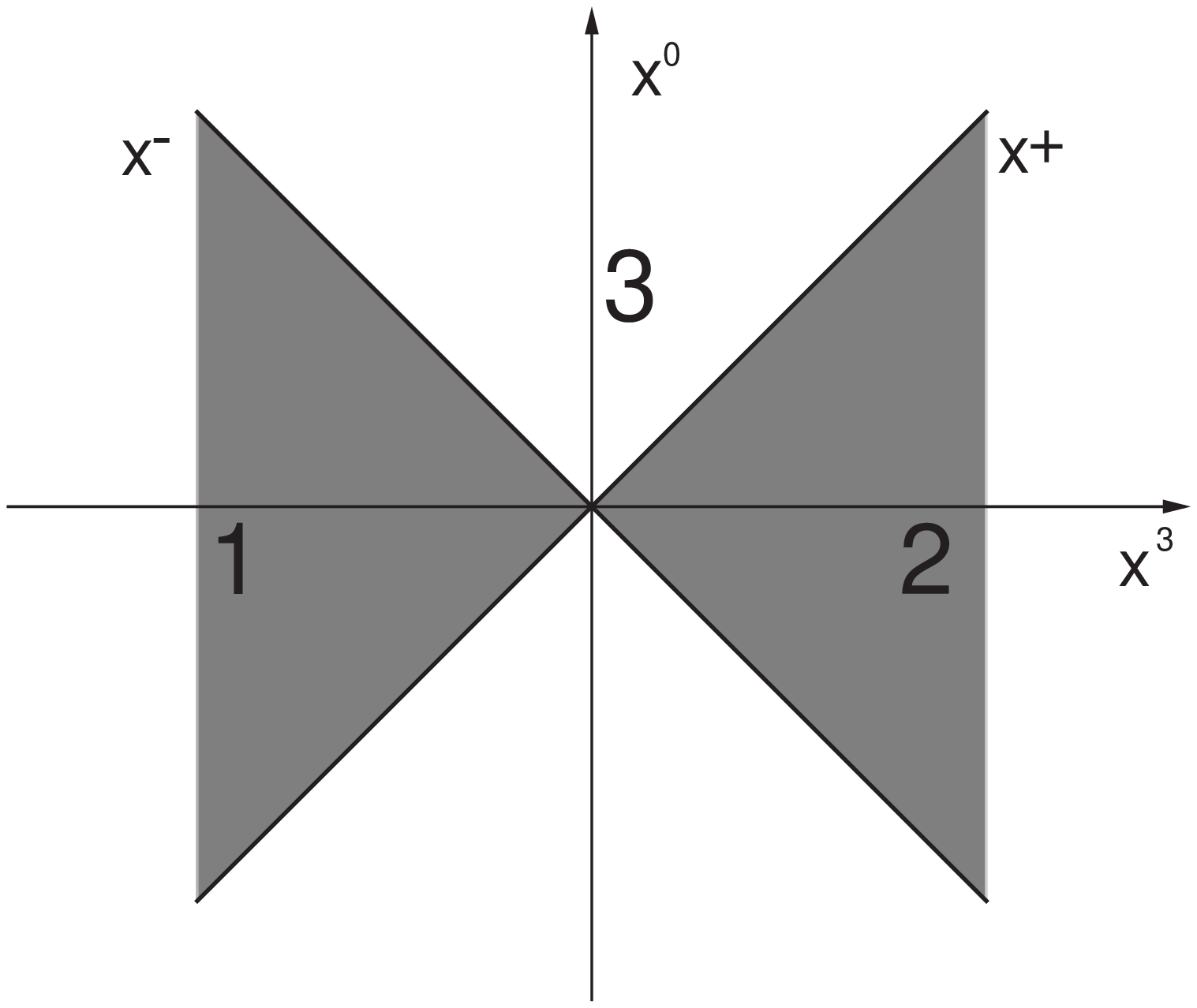}
	\end{minipage} \hfill
	\begin{minipage}[b]{\medcapsize} \caption[] {
		\label{regions} \sloppy \small Regions with different
		structures of the gauge potential: \\
	       	In regions 1 and 2 we have the well known one nucleus
		solutions $\alpha_{1,2}$. While the gauge potential
		in the backward light cone is vanishing we have a
		nontrivial solution in the forward lightcone,
		region 3
		}
	\vspace{1 truecm} \end{minipage}
\end{figure}

Let us now turn to nucleus-nucleus collisions. Obviously
the single nucleus solutions are still
valid everywhere except in those regions of space-time which are in
causal contact with the collision point, i.e. in its forward light
cone (see Fig. \ref{regions}).
In the abelian case even there the solutions would be directly given
in terms of the single nucleus solutions as a sum of two pure gauge
fields which again constitutes a pure gauge field. Hence in this case
there would be {\em no} nontrivial effects on the classical field
level. To generate photons and field strength one would have to resort
to quantum effects. This is fundamentally different in the non-Abelian
case.  Obviously now, the sum of two pure gauge field is no longer a
pure gauge field due to nonlinear effects, and therefore it does not
solve the Yang Mills equations for the two nuclei problem.
Nevertheless the one nucleus solutions provide initial conditions for
the evolution of the gauge field in the forward light cone as
discussed in \cite{komcwe9502}.

In the remainder of this paper we will adopt
the gauge condition
\begin{eqnarray}
x^+A^- + x^-A^+ = 0 \label{radgauge}
\end{eqnarray}
In this gauge the fields in the
forward light cone can be cast in the form
\begin{eqnarray}
A^\pm & = & \pm x^\pm \alpha(\tau,x_\perp)
\nonumber \\
A^i & = & \alpha_\perp^i(\tau,x_\perp)
	\label{Afwlc}
\end{eqnarray}
where $\tau =
\sqrt{2 x^+ x^-}$ is the proper time.
Note that this representation does not introduce any physical
assumption not already present in the single nucleus solution.  The
relation between $A^\pm$ is a consequence of the gauge
condition. Further, the fact that the functions $\alpha(\tau,x_\perp)$
and $\alpha_\perp^i(\tau,x_\perp)$
do not depend on the space time rapidity
$\eta = 1/2 \ln{x^+/x^-}$ is a natural consequence of the absence of a
longitudinal length scale in our initial conditions with its infinitely
thin nuclei moving at the speed of light.

Since the valence color charge densities vanish inside the light cone,
the fields $A_\mu$ of eq. (\ref{Afwlc}) should solve the homogeneous
equations
\begin{eqnarray}
\left[D_\mu,F^{\mu\nu}\right] & = & 0 \label{YMfwlc}
\end{eqnarray}
The initial
conditions are determined by patching the solutions in the three
regions (see Fig. \ref{regions})
together and eliminating the discontinuities at $\tau=0$.
This specifies the forward light cone fields
$\left.\alpha^i_\perp\right\vert_{\tau = 0}$ and
$\left.\alpha\right\vert_{\tau = 0}$ in terms of the known single
nucleus solutions outside the forward light cone.

\section{Perturbation theory}

Using the form \mref{Afwlc} of the gauge potential, the Yang-Mills
equations \mref{YMfwlc} in the forward light cone are
\begin{eqnarray}
{1\over\tau^3}\partial_\tau \tau^3 \partial_\tau\alpha
	+ \left[ D_i, \left[D^i,\alpha\right] \right]
	& = & 0
\nonumber \\
{1\over\tau} \left[ D_i,
	\partial_\tau \alpha_\perp^i \right]  +
	ig \tau \left[\alpha,\partial_\tau\alpha\right] & = & 0
\nonumber \\
{1\over \tau}\partial_\tau \tau \partial_\tau\alpha^i_\perp
	- ig \tau^2\left[
	\alpha, \left[D^i,\alpha\right]
	\right] - \left[D^j,F^{ji}\right] & = & 0 \label{eqofm}
\end{eqnarray}
Matching conditions on the solution lead to the initial conditions for
$\alpha(\tau,x_\perp)$ and $\alpha_\perp^i(\tau,x_\perp)$ in terms of
the single nucleus solutions $\alpha_m^i$:
\begin{eqnarray}
\left.\alpha^i_\perp\right\vert_{\tau = 0} & = &
	\alpha_1^i + \alpha_2^i
\nonumber \\
\left.\alpha\right\vert_{\tau = 0} & = &
	{i g\over 2}\left[\alpha_1^i,\alpha_2^i\right]
\label{eq:init}
\end{eqnarray}

In this paper we construct solutions in the weak field limit by
expanding first the initial conditions and then the fields within the
forward light cone in powers of $\rho$.

Let us first concentrate on the initial conditions.  To determine the
initial conditions perturbatively, we need the single nucleus fields
$\alpha_m^i$, $i=1,2$ to the appropriate order.

Recall that for a single nucleus, the Yang-Mills equations reduce to
the two equations \mref{eq:sing}. The first states that $\alpha_m^i$
is a pure gauge field. The second equation determines $\alpha^i_m$ in
terms of the charge density $\rho_m(x_\perp)$ of the individual nuclei
via
\begin{eqnarray}
\partial^i \alpha_m^i \ \ = \ \
	\partial^i \left\{-{1\over i g}
	U_m(x_\perp)\partial^i U_m^{-1}\right\}
	& = & g \rho_m(x_\perp)
	\label{YMonenuc}
\end{eqnarray}
To second order in $\rho$, the solution of \mref{YMonenuc} is
\begin{eqnarray}
{\alpha}_m^i & = &
  -\,\partial^{i}\phi_m
%\nonumber \\ &&
+ 	{i g\over 2}
	\left(\delta^{ij}
	-\partial^i{1\over \np^2}\partial^j\right)
	\left\{\left[
	\phi_m,
	\partial^{j}\phi_m
           \right] + {\rm O}(\rho_m^3)\right\}
\label{eq:alphak}
\end{eqnarray}
where we have defined
\begin{eqnarray}
\phi_m & = & -{g\over \np^2} \rho_m
\label{eq:phidef}
\end{eqnarray}
Note that the first order term in \mref{eq:alphak} is longitudinal,
whereas all higher corrections are transverse. Their sole purpose is
to render $\alpha^i_m$ a pure gauge field. Plugging \mref{eq:alphak}
into \mref{eq:init} gives us initial conditions up to second order in
$\rho_m$.

To solve the equations of motion \mref{eqofm} in the forward light
cone perturbatively, we also have to expand the gauge fields there in
powers of $\rho_m$
\begin{eqnarray}
\alpha & = & \sum\limits_{n=0}^\infty  \alpha_{(n)}
\nonumber \\
\alpha^i_\perp & = & \sum\limits_{n=0}^\infty
	 {\alpha^i_\perp}_{(n)}
\end{eqnarray}

\subsubsection*{First order:}

To first order the equations of motion are necessarily linear
\begin{eqnarray}
 {1\over \tau^3} \partial_\tau \tau^3
	\partial_\tau \alpha_{(1)} -
	\np^2 \alpha_{(1)}
	& = & 0
\nonumber \\
 \partial^{i}\partial_\tau {\alpha_\perp}_{(1)}^{i} & = & 0
\nonumber  \\
 {1\over \tau} \partial_\tau \tau \partial_\tau
	{\alpha_\perp}_{(1)}^{i}
	-\left(\np^2 \delta^{ij} - \partial^i\partial^j\right)
	{\alpha_\perp}_{(1)}^{j}
	& = & 0
\end{eqnarray}
They are to be solved subject to the initial conditions
\begin{eqnarray}
\left.{\alpha}_{(1)}(\tau,x_\perp)\right.\vert_{\tau=0}
& = & 0
\nonumber \\
\left.{\alpha_\perp^i}_{(1)}(\tau,x_\perp)\right.\vert_{\tau=0}
& = & -\, \partial^i\left(
	{\phi^1} +{\phi^2}\right)(x_\perp)
\end{eqnarray}
 The initial
condition for the ``$\pm$''-components to this order is trivial.

With these initial conditions, the first order solutions are
completely specified and essentially trivial. We obtain vanishing
$\alpha_{(1)}$
\begin{eqnarray}
\alpha_{(1)}(\tau,x_\perp) & = & 0
\end{eqnarray}
and $\tau$-independent ${\alpha_\perp^i}_{(1)}$
\begin{eqnarray}
{\alpha_\perp^i}_{(1)}(\tau,x_\perp) & = &
-\, \partial^i\left(
	{\phi^1} +{\phi^2}\right)(x_\perp)
\end{eqnarray}
which obviously is {\em pure gauge}.

To this order the problem is structurally identical to its abelian
counterpart. Commutators do not appear neither in the equations of
motion nor in the
initial conditions (these contributions are necessarily
at least second order in $\rho_m$), and hence no field strength
is generated in the forward light cone.

\subsubsection*{Second order:}

In the second order this changes immediately. Using the first order
information almost all nonlinear terms drop from the second order
equations which now read
\begin{eqnarray}
 {1\over \tau^3} \partial_\tau \tau^3 \partial_\tau \alpha_{(2)}
	-\np^2 \alpha_{(2)} & = & 0
\nonumber \\
	\partial^{i}\partial_\tau{\alpha_\perp}_{(2)}^{i} & = & 0
\nonumber \\
{1\over \tau} \partial_\tau \tau \partial_\tau
	{\alpha_\perp}_{(2)}^{i}
	- \left(\np^2 \delta^{ij} - \partial^i\partial^j\right)
	{\alpha_\perp}_{(2)}^{j}
+ i g\, \partial^{j}[{\alpha_\perp}_{(1)}^{j},
	{\alpha_\perp}_{(1)}^{i}]
              & = & 0
\end{eqnarray}
The only inhomogeneity in these equations comes from the -- pure gauge
-- ${\alpha_\perp}^i_{(1)}$.  In fact using the residual
($\tau$-independent) gauge freedom, we can remove the inhomogeneous
piece altogether.

We define $\epsilon_\mu$ by
\begin{eqnarray}
A_\mu & = & V(x_\perp)
	\left[ \epsilon_\mu - {1\over i g} \partial_\mu\right]
	V^{-1}(x_\perp)
\label{eq:epsdef}
\end{eqnarray}
where $V(x_\perp) = \exp -ig \phi_\perp(x_\perp)$.
$\phi_\perp(x_\perp)$ is to be determined such that it removes
nonlinear terms from the equations of motion for $\epsilon$
at least to second order.
In fact, it turns out, that it is possible to remove
the inhomogeneous terms up to third order in $\rho$
requiring
\begin{eqnarray}
\partial^i\left.{\epsilon^i}\right.\vert_{\tau = 0} & = & 0
\label{eq:resgauge}
\end{eqnarray}
In higher orders it is no longer
possible to linearize the equations.
Using equations \mref{eq:epsdef} and \mref{eq:resgauge} we
find for ${\phi_\perp}$
\begin{eqnarray}
{\phi_\perp} = \phi_1 + \phi_2 + {\rm O}(\rho^2)
\end{eqnarray}
With this choice of $V$ the first order solution vanishes
\begin{eqnarray}
\epsilon_{(1)} & \equiv & 0
\nonumber \\
\epsilon^i_{(1)} & \equiv & 0
\end{eqnarray}

The second order equations become
\begin{eqnarray}
  {1\over \tau^3} \partial_\tau \tau^3 \partial_\tau
\epsilon_{(2)} -\np^2 \epsilon_{(2)}
  & = & 0
\nonumber \\
  	\partial^{i}\partial_\tau{\epsilon}_{(2)}^{i}
        & = & 0
\nonumber \\
{1\over \tau} \partial_\tau \tau \partial_\tau
	{\epsilon}_{(2)}^{i}
	-\left(\delta^{ij} \np^2
		- \partial^i\partial^j\right)
	{\epsilon}_{(2)}^{j}
	 & = & 0
\end{eqnarray}
The initial conditions on $\alpha_\perp^i$ and $\alpha$ are
transformed into
\begin{eqnarray}
\left.\epsilon^i_{(2)}\right.\vert_{\tau = 0} & = &
	-i g\
	\left(\delta^{ij}
	-\partial^i{1\over \np^2}\partial^j\right)
	\left[\phi_{1},
	\partial^{j}\phi_{2}
           \right]
\end{eqnarray}
\begin{eqnarray}
\left.\epsilon_{(2)}\right.\vert_{\tau = 0} & = &
{i g\over 2}\left[\partial^i{\phi_1},
		\partial^i{\phi_2}\right]
\end{eqnarray}
Obviously, the gauge condition
$\partial^i\left.{\epsilon^i}\right.\vert_{\tau = 0} = 0$ in this
order is preserved for all proper times $\tau$. Hence the
``$i$''-components of the gauge field may be represented as
\begin{eqnarray}
\epsilon^i & = & \epsilon^{ij}\partial^j\chi
\end{eqnarray}
Using $\epsilon$ and $\chi$, the second order Yang-Mills equations and
initial conditions simplify even further. In terms of these variables,
we finally obtain
\begin{eqnarray}
  {1\over \tau^3} \partial_\tau \tau^3 \partial_\tau
\epsilon_{(2)} -\np^2 \epsilon_{(2)}
  & = & 0
\nonumber \\
{1\over \tau} \partial_\tau \tau \partial_\tau
	{\chi}_{(2)}
	-\np^2 {\chi}_{(2)}
	 & = & 0
\label{bessels}
\end{eqnarray}
and
\begin{eqnarray}
\left.\epsilon_{(2)}\right.\vert_{\tau = 0} & = &
{i g\over 2}\left[\partial^i{\phi_1},
		\partial^i{\phi_2}\right]
%\ \ = \ \ {i g\over 2}\left[{\alpha_1}^i,
%		{\alpha_2}^i \right]
\nonumber \\
\left.\chi_{(2)}\right.\vert_{\tau = 0} & = & -ig\
	\epsilon^{ij}\
	\left[\partial^i{\phi_1},
		\partial^j{\phi_2}\right]
%\ \ = \ \ -ig\
%	\epsilon^{ij}\
%	\left[{\alpha_1}_{(1)}^i,
%		{\alpha_2}_{(1)}^j \right]
\label{initialconds}
\end{eqnarray}
We stress that, even though the equations of motion \mref{bessels}
linearize to this order due to our gauge choice, the intrinsic
nonlinearities of the system are very important. They show up in the
initial conditions \mref{initialconds} which carry nontrivial physical
information.

\section{Perturbative solutions}

To solve \mref{bessels}, we note that the equations are of form
\begin{eqnarray}
{1\over \tau^m} \partial_\tau \tau^m \partial_\tau
f(\tau,x_\perp) -\np^2 f(\tau,x_\perp)
\end{eqnarray}
where in our case $m = 1,3$, $f = \chi,\epsilon$. Factorizing $\tau$
and $x_\perp$-dependence, we immediately identify the eigenfunctions
\begin{eqnarray}
f_{k_\perp}(\tau,x_\perp) & = &
	(\omega_k\tau)^{1-m\over 2} \
	Z_{\pm\left\vert{m-1\over 2}\right\vert}
	(\omega_k \tau) \ e^{i k_\perp^j x_\perp^j}
\end{eqnarray}
where $\omega_k = \sqrt{k_\perp^2}$ and $Z_\nu$ may in general be any
linear combination of the Bessel functions $J_\nu,N_\nu$.

The initial conditions \mref{initialconds} force our solutions to be
regular at $\tau = 0 $ and thus preclude any admixture of
Neumann-functions $N_\nu$.

A general solution is therefore of the form
\begin{eqnarray}
F_m(\tau,x_\perp) & = & \int\!\!{d^2\!k_\perp\over (2\pi)^2}
	\ h_m(k_\perp)\  f_{k_\perp}(\tau,x_\perp)
\nonumber \\ & = &
	\int\!\!{d^2\!k_\perp d^2\!y_\perp \over (2\pi)^2}
	\
	e^{i k_\perp^j (x_\perp-y_\perp)^j}
	\
	h_m(y_\perp)
	\
	(\omega_k \tau)^{1-m\over 2} \
	J_{\left\vert{m-1\over 2}\right\vert}
	(\omega \tau)
\end{eqnarray}
For  equations \mref{bessels} this yields
\begin{eqnarray}
\epsilon_{(2)}(\tau, x_\perp) & = & \int\!\!
	{d^2\!k_\perp d^2\!y_\perp\over (2\pi)^2}
	e^{i k_\perp^j(x-y)_\perp^j}
	\
	h_3(y_\perp)
	\
	{1\over \omega\tau} J_1(\omega \tau)
\nonumber \\
\chi_{(2)}(\tau, x_\perp) & = & \int\!\!
	{d^2\!k_\perp d^2\!y_\perp\over (2\pi)^2}
	\
	e^{i k_\perp^j(x-y)_\perp^j}
	\
	h_1(y_\perp) J_0(\omega \tau)
\end{eqnarray}
with $h_3$ and $h_1$ determined by the initial conditions
eq.(\ref{initialconds}) as
\begin{eqnarray}
	 h_3(x_\perp) & = &
	{i g}
	\left[\partial^i \phi_1,
	\partial^i \phi_2\right](x_\perp)
\nonumber \\
h_1(x_\perp) & = &
	- i g {\epsilon^{i j}\over \np^2}
	\left[\partial^i\phi_1,
	\partial^j\phi_2\right](x_\perp)
\end{eqnarray}
In the following we will need the large $\tau$ asymptotics of the
solutions:
\begin{eqnarray}
\chi_{(2)}(\tau, x_\perp) & = & \int\!\!
	{d^2\!k_\perp \perp\over (2\pi)^2}
	\
	e^{i k_\perp^j x_\perp^j}
	\
	h_1(k_\perp)
	\
	\left\{
	{\sqrt{2/\pi}\over (\omega\tau)^{1/2}}
		\cos(\omega\tau -{\pi\over 4}) +
		{\rm O}({1\over \tau^{3/2}})
	\right\}
	\label{iasymp}
\end{eqnarray}
\begin{eqnarray}
\epsilon_{(2)}(\tau, x_\perp) & = & \int\!\!
	{d^2\!k_\perp\over (2\pi)^2}
	e^{i k_\perp^j x_\perp^j}
	\
	h_3(k_\perp)
	\
	\left\{
		{\sqrt{2/\pi} \over (\omega\tau)^{3/2}}
		\cos(\omega\tau -{3\over 4}\pi) +
		{\rm O}({1\over \tau^{3/2}})
	\right\}
	\label{pmasymp}
\end{eqnarray}

\section{Distribution functions \label{sec:dist}}

The solutions of the Yang Mills equations at asymptotically large $\tau$
are of the form
\begin{eqnarray}
	\epsilon^a(\tau,x_\perp) & = &
      \int {{d^2k_\perp} \over {(2\pi)^2}}
       {1 \over \sqrt{2\omega}}
      \left\{ a_1^a(\vec{k}_\perp) {1 \over \tau^{3/2}}
      e^{ik_\perp\cdot x_\perp -i\omega \tau} + C. C.
        \right\}
\nonumber \\
        {\epsilon}^{a,i} (\tau,x_\perp) & = &
	\int {{d^2k_\perp} \over {(2\pi)^2}}
        \kappa^i
{1 \over \sqrt{2\omega}} \left\{ a_2^a(k_\perp)
{1 \over \tau^{1/2}}
      e^{ik_\perp x_\perp-i\omega \tau} + C. C. \right\}
\end{eqnarray}
Here $\omega = \mid k_\perp \mid$, and $\kappa^i = \epsilon^{ij} k^j/\omega$.
The notation $C. C.$ means complex conjugate.

To derive an expression for the energy density, we recall that $\tau $
is large.  Near $z = 0$, this implies that in the range of $z$ where
$\tau \sim t \gg z$ the solutions are $z$ independent.  This means
they asymptotically have zero $p_z$.  Now suppose we are at any value
of $z$, and $\tau$ is large but $t \sim z$.  We can do a longitudinal
boost to $z = 0$ without changing the solution.  Again in this frame
the solution has zero $p_z$.  We see therefore that for the asymptotic
solutions the space time rapidity is one to one correlated with
the momentum space rapidity, that is at asymptotic times we find that
\begin{eqnarray}
	\eta = {1 \over 2} \ln(x^+/x^-)
	= y = {1 \over 2} \ln(p^+/p^-)
\end{eqnarray}

To proceed further, we compute the energy density in the neighborhood
of $z = 0$.  Here asymptotically $\tau = t$.  The energy in a box of
size $R$ in the transverse direction and $dz$ in the longitudinal
direction, with $L \ll t$ becomes \cite{blaizot}
\begin{eqnarray}
	dE = {{dz} \over t} \int {{d^2k_\perp}\over {(2\pi)^2}}
	\omega \sum_{i,b} \mid a_i^b(k_\perp) \mid^2
\end{eqnarray}
Recalling that $dy = dz/t$, we find that
\begin{eqnarray} {{dE} \over {dyd^2k_\perp}} =
{1 \over {(2\pi)^3}} \omega \sum_{i,b} \mid a_i^b(k_\perp) \mid^2
\end{eqnarray}
and the multiplicity distribution of gluons is
\begin{eqnarray}
{{dN} \over {dyd^2k_\perp}} & = &
	 {1 \over \omega} {{dE} \over {dyd^2k_\perp}}
\end{eqnarray}
As we expect for a boost covariant solution, the multiplicity
distribution is rapidity invariant.

This is now easily compared to the asymptotic behavior of the second
order solutions \mref{pmasymp}, \mref{iasymp}. We read off that
\begin{eqnarray}
a_1(k_\perp) & = & {1\over\sqrt\pi} {h_3(k_\perp)\over \omega}
%\nonumber \\ & = & {1\over\sqrt\pi} \int\!\!{d^2\!y_\perp}
%			e^{-ik_\perp{\cdot}x_\perp}
%			{h_3(y_\perp)\over \omega}
\nonumber \\
a_2(k_\perp) & = &  {1\over\sqrt\pi} i\omega h_1(k_\perp)
%\nonumber \\ & = &  {1\over\sqrt\pi} i\omega
%			\int\!\!{d^2\!y_\perp}
%			e^{-ik_\perp{\cdot}x_\perp}
%			h_1(y_\perp)
\end{eqnarray}
Hence we get the distribution function
\begin{eqnarray}
{{dN} \over {dyd^2k_\perp}} & = &
	 {1 \over \omega} {{dE} \over {dyd^2k_\perp}}
\nonumber \\ & = &
	{1 \over {(2\pi)^3}} \sum_{i}
	2 \ \tr \ a_i(k_\perp) a_i^\dagger
\nonumber \\ & = &
	{1 \over {(2\pi)^3}} \sum_{i}
	{2\over\pi} \ \tr \left\{
	{h_3(k_\perp)h_3^\dagger(k_\perp)\over \omega^2}
	+
	\omega^2 h_1(k_\perp)h_1^\dagger(k_\perp)
	\right\}
\end{eqnarray}
Now we use that
\begin{eqnarray}
h_3(k_\perp) & = &
	\int\!\!d^2\!y_\perp e^{-ik_\perp{\cdot}y_\perp}
	\ i g\left\{
	\left[\partial^i\phi_1,
		\partial^i\phi_2\right]
	\right\}(y_\perp)
\nonumber \\
h_1(k_\perp) & = &
	\int\!\!d^2\!y_\perp e^{-ik_\perp{\cdot}y_\perp}
	\ \left\{-i g {\epsilon^{i j}\over \np^2}
		\left[\partial^i\phi_1,
		\partial^j\phi_2\right]
	\right\}(y_\perp)
\nonumber \\ & = &
	{1\over \omega^2}
	\int\!\!d^2\!y_\perp e^{-ik_\perp{\cdot}y_\perp}
	\ \left\{i g \epsilon^{i j}
		\left[\partial^i\phi_1,
		\partial^j\phi_2\right]
	\right\}(y_\perp)
\end{eqnarray}
and get
\begin{eqnarray}
{{dN} \over {dyd^2k_\perp}} & = &
	{\rm S}_\perp {g^2 \over {(2\pi)^3}}
	{2\over\pi k_\perp^2}
	\left[ \delta^{i j}\delta^{k l}
		+ \epsilon^{i j}\epsilon^{k l}\right]
	\int\!\!d^2\!x_\perp
	e^{ik_\perp{\cdot}x_\perp}
\nonumber \\ &&
	\quad \tr
	\left\langle
	\left[\partial^i\phi_1,
		\partial^j\phi_2\right](x_\perp)
	\left[\partial^k\phi_1,
		\partial^l\phi_2\right]^\dagger(0)
	\right\rangle_\rho
\end{eqnarray}
where, S$_\perp$ is the transverse area of the system. As indicated,
we have to average over $\rho$ with the Gaussian weight
eq.(\mref{eq:rhoav}).  Using \mref{eq:phidef} we find
\begin{eqnarray}
\langle \phi_m^a(x_\perp) \phi_n^b(y_\perp) \rangle_\rho & = &
	g^2\,\mu^2 \ \delta^{m n}\ \delta^{a b}\
		\int\!\!{d^2\!p_\perp\over (2\pi)^2}
		{e^{ip_\perp{\cdot}(x-y)_\perp}
		  \over p_\perp^4}
\label{eq:phiprops}
\end{eqnarray}
and the above turns into
\begin{eqnarray}
{{dN} \over {dyd^2k_\perp}} & = &
	{\rm S}_\perp {g^6 \over {(2\pi)^3}}
	{2\over\pi k_\perp^2}
	\left[ \delta^{i j}\delta^{k l}
		+ \epsilon^{i j}\epsilon^{k l}\right]
\nonumber \\ &&
	\mu^4 \tr\left([\tau^a,\tau^b][\tau^b,\tau^a]\right)
	\int\!\!{d^2\!p_\perp\over (2\pi)^2}
	{p^i p^k (p+k)^j (p+k)^l\over p^4 (p+k)^4}
\nonumber \\ & = &
	{\rm S}_\perp {g^6 \over {(2\pi)^3}}
	{2\over\pi k_\perp^2}
	\mu^4 N_c(N_c^2-1)
	\int\!\!{d^2\!p_\perp\over (2\pi)^2}
	{1\over p^2 (p+k)^2}
\label{eq:distloop}
\end{eqnarray}
The integral in equation \mref{eq:distloop} is infrared
divergent. This however is an artifact of the weak field expansion
employed above. As discussed in \cite{mv1a,mv1b,mv2,komcwe9502}, the
inclusion of higher order terms for the initial fields would generate a
mass scale of order $\alpha_s \mu$ and hence regulate this
divergence. With logarithmic accuracy we obtain therefore
\begin{eqnarray}
{{dN} \over {dyd^2k_\perp}} & = &
	{\rm S}_\perp {g^6 \over {(2\pi)^3}}
	{2\over\pi}
	\mu^4 N_c(N_c^2-1)
	\ \
	{1\over 2\pi} {1\over k_\perp^4}
	\ln{k_\perp^2\over (\alpha_s \mu)^2}
\label{dN}
\end{eqnarray}
This is the main result of this work.

\section{Conclusion}

In this paper we have calculated the high $k_\perp$ asymptotics of the
distribution function of gluons produced in a collision of two
ultrarelativistic heavy nuclei in the framework of the
McLerran-Venugopalan model.

Our result, eq. (\ref{dN}) has a very natural interpretation in
analogy to the naive parton model. The number of produced gluons in
the parton model is given by
\begin{eqnarray}
N(k_\perp)=\int d^2 p_\perp N_1(p_\perp)
	N_2(k_\perp-p_\perp)\sigma(p_\perp,k_\perp)
\label{parton}
\end{eqnarray}
where $N_m(q_\perp)$ is the probability to find a gluon with
transverse momentum $q_\perp$ in the $m$-th nucleus, and $\sigma$ is
the cross section for production of the gluon with momentum $k_\perp$
in the final state in the collision of the two gluons.  To the lowest
order in perturbation theory the cross section is
\begin{eqnarray}
\sigma(k_\perp)\propto \frac{\alpha_s}{k_\perp^2}
\end{eqnarray}
The distribution functions for the one nucleus case $N_m$ were
calculated in \cite{mv1a,mv1b} and were found to have the same
functional form as in the lowest order in perturbation theory
\begin{eqnarray}
N_m(q_\perp)\propto \frac{\alpha_s\mu^2}{q_\perp^2}
\end{eqnarray}
Substituting this into eq.(\ref{parton}) gives the result of
eq.(\ref{dN}).

\begin{figure}[htb]
	\hfill
	\begin{minipage}[b]{\medfigsize} \epsfxsize 3.5cm
		\epsfbox{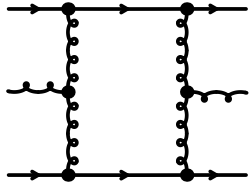}
	\end{minipage} \hfill
	\begin{minipage}[b]{\medcapsize}
	\caption[] {
		\small
		Diagrammatic representation of the lowest
		order distribution function
		\label{fig:distdiagram}
	 }
	\vspace{0.5 truecm}
	\end{minipage}
\end{figure}

Indeed the calculation performed in this paper can be represented in
terms of Feynman diagrams. The gluon distribution function of
eq.(\ref{dN}) has a representation of Fig. \ref{fig:distdiagram}.

The reader may have noticed one peculiarity in the preceding
discussion. Although we have been using the parton model notations,
some of the gluons involved in the process depicted on
Fig. \ref{fig:distdiagram} are virtual. To produce the real final
state gluon, the internal gluons must be off shell with virtuality of
order of their momentum. In fact if one wants to think
about this process in terms of real partons,
it must be represented as ``two quarks go into two
quarks and a gluon'', rather than ``two gluons go into one gluon'',
which is kinematically forbidden for on shell particles.

As we have seen, to this order our calculation essentially reproduces
perturbation theory. The reason is that we have expanded in powers of
the valence charge density, and thereby have ignored the fact that the
classical fields involved are expected to be strong.  The real
nonperturbative nature of this approach will become apparent when
those are fully taken into account.  Some of the qualitative
consequences of such an analysis however, can be understood already at
this point.  Consider for example, the total multiplicity of produced
gluons.
\begin{eqnarray}
N=\int d^2k_\perp {{dN} \over {dyd^2k_\perp}}
\end{eqnarray}
Since ${{dN} \over {dyd^2k_\perp}}$ is proportional to the transverse
area of the system, and is a function of the dimensionless ratio
$\mu^2/k_\perp^2$, the integral has to be proportional to
$S_\perp\mu^2$.  Remembering that for large nuclei $\mu$ scales with
the number of nucleon in the nucleus $A$ as $\mu^2\propto A^{1/3}$, we
conclude that for large nuclei the total multiplicity scales as
\begin{eqnarray}
N\propto A
\end{eqnarray}
On the other hand, the total multiplicity at momenta larger than some
fixed momentum $p\gg \alpha\mu$ will coincides with the perturbative
result
\begin{eqnarray}
N=\int\limits_{k_\perp^2\geq p^2} d^2k_\perp {{dN} \over {dyd^2k_\perp}}
\propto S_\perp\mu^4\propto A^{4/3}
\end{eqnarray}

Therefore it is very important to go beyond the weak field expansion
and incorporate truly nonperturbative effects due to strong
classical fields.

In comparing our results with the parton cascade model of Geiger et al
\cite{gemu92,gemu93,gemu94} one should keep in mind the following
differences between the two approaches.  In the parton cascade, the
leading order in $\alpha_s$ scattering occurs by gluon gluon
scattering.  This populates the high $p_\perp$ tail of the
distribution by scatterings of low $p_\perp$ gluons.  In our case, we
have an intrinsic $p_\perp$ for the gluons and the gluons are far
enough off mass shell so that we can produce high $p_\perp$ gluons by
glue-glue goes to single gluon scattering.  Consequently, although the
$p_\perp$ dependence of the two results is similar, in the high
$p_\perp$ region we have a different dependence on $\alpha_S$.  We
have one power less, since the factor $\alpha^2\mu^2$ in
eq. (\ref{dN}) just sets the scale of the intrinsic single nucleus
glue distribution.

The scattering processes, which are leading in our approximation are
also present in cascade codes as composite processes $qq\rightarrow
qqg$, where some of the quarks are off shell by an amount $p_\perp^2$.
Although these contributions are formally of order $\alpha^3$, for
very large nuclei the quark distribution functions will be large and
these processes may become leading. This seems to be the natural point
of contact between the two approaches.

Of course at some $p_\perp \gg \mu$, our approximations for computing the
gluon distribution function break down.  This presumably first occurs
when gluon brems\-strahlung softens the high $p_\perp$ distribution from its
$1/p_\perp^2$ behavior, and the spectrum steepens.
The main point is however not what happens in the tail of the
distribution.  This is simply where it is most easy to compute.  In
the center of the $p_\perp$ distribution of produced gluons, the gluons
arise from a non-linear evolution of the gluons fields.  This is cause
by the quantum mechanical nature of the initial state and the charge
coherence of the initial state interactions.  In this case, the final
state distribution of gluons bears scant resemblance to that of the
initial distribution convoluted over hard gluon scattering.  Before one
knows whether there are substantial quantitative differences, one must
of course numerically solve the problem for gluons in the center of
the distribution.

\section*{Acknowledgments}

This research was supported by the U.S. Department of Energy under
grants No. DOE High Energy DE--AC02--83ER40105 and No. DOE Nuclear
DE--FG02--87ER--40328. One of us (HW) was supported by a fellowship of
the Alexander van Humboldt Foundation.

\newcommand{\noopsort}[1]{}


\begin{thebibliography}{1}

\bibitem{mv1a}
L.~McLerran and R.~Venugopalan.
\newblock {\em Phys. Rev.}, {\bf D49} (1994), 2233.

\bibitem{komcwe9502}
A.~Kovner, L.~{McLerran}, and H.~Weigert.
\newblock {\em Gluon Production from Non-Abelian Weizs\"acker-Williams Fields
  in Nucleus-Nucleus Collisions}.
\newblock {\sl submitted to Phys. Rev. D} {\bf hep-ph/9502289}, 1995.

\bibitem{mu94}
A.~Mueller.
\newblock {\em Nucl. Phys.}, {\bf B425} (1994), 373.

\bibitem{mv1b}
L.~McLerran and R.~Venugopalan.
\newblock {\em Phys. Rev.}, {\bf D49} (1994), 3352.

\bibitem{mv2}
L.~{McLerran} and R.~Venugopalan.
\newblock {\em Phys. Rev.}, {\bf D50} (1994), 2225.

\bibitem{blaizot}
J.~P. Blaizot and A.~Mueller.
\newblock {\em Nucl. Phys.}, {\bf B289} (1987), 847.

\bibitem{gemu92}
K.~Geiger and B.~M\"uller.
\newblock {\em Nucl. Phys.}, {\bf B369} (1992), 600.

\bibitem{gemu93}
K.~Geiger.
\newblock {\em Phys. Rev.}, {\bf D47} (1993), 133.

\bibitem{gemu94}
K.~Geiger and B.~M\"uller.
\newblock {\em Cern Preprint} {\bf CERN-TH 7313/94}, 1994.

\end{thebibliography}
\end{document}